\documentclass[twocolumn,aps,showpacs]{revtex4-2}
\usepackage[latin9]{inputenc}
\usepackage{textcomp}
\usepackage{amsmath}
\usepackage{amssymb}
\usepackage{graphicx}
\usepackage{subscript}

\makeatletter

\DeclareFontEncoding{LGR}{}{}
\DeclareRobustCommand{\greektext}{%
  \fontencoding{LGR}\selectfont\def\encodingdefault{LGR}}
\DeclareRobustCommand{\textgreek}[1]{\leavevmode{\greektext #1}}
\ProvideTextCommand{\~}{LGR}[1]{\char126#1}

\providecommand{\tabularnewline}{\\}

\usepackage{color}

\def\NOT(#1,#2){\OneQubitGate(#1,#2){$X$}}

\makeatother

\begin{document}
\title{Characterization of single shallow silicon-vacancy centers in $4H$-SiC}
\author{Harpreet Singh\textsuperscript{1,2}, Mario Alex Hollberg\textsuperscript{1},
Misagh Ghezellou\textsuperscript{3}, Jawad Ul-Hassan\textsuperscript{3},
Florian Kaiser\textsuperscript{4} and Dieter Suter\textsuperscript{1}\\
 \textsuperscript{1}Fakultät Physik, Technische Universität Dortmund,
Dortmund, Germany. \textsuperscript{2}Department of Chemistry, University
of California, Berkeley, Berkeley, CA 94720, USA. \textsuperscript{3}Department
of Physics, Chemistry and Biology, Linköping University, Linköping,
Sweden. \textsuperscript{4}3rd Institute of Physics and Research
Center SCOPE, University of Stuttgart, Stuttgart, Germany.}
\begin{abstract}
Shallow negatively charged silicon-vacancy centers have applications
in magnetic quantum sensing and other quantum applications. Vacancy
centers near the surface (within 100 nm) have different spin relaxation
rates and optical spin polarization, affecting the optically detected
magnetic resonance (ODMR) signal. This makes it essential to characterize
these centers. Here we present the relevant spin properties of such
centers. ODMR with a contrast of up to 6 \%, which is better than
the state of the art, allowed us to determine the zero field splitting,
which is relevant for most sensing applications. We also present intensity-correlation
data to verify that the signal originates from a single center and
to extract transition rates between different electronic states.
\end{abstract}
\maketitle

\section{Introduction}

Silicon vacancies in silicon carbide (SiC) are excellent candidates
for quantum sensing and information applications$\;$\cite{koehl-nature-11,kraus-nature-13,widmann-nature-14,nagy2019high}.
Negatively charged silicon vacancies ($V_{Si}^{-}$) have spin 3/2$\;$\cite{soltamov-prl-12,riedel-prl-12}.
The spin of $V_{Si}^{-}$ can be initialized into a particular spin
groundstate by optical pumping$\;$\cite{soltamov-prl-12,nagy2019high,singh-prb-21}
and it has sufficiently long coherence and relaxation times for many
interesting applications$\;$\cite{widmann-nature-14,carter-prb-15,simin-prb-17,kraus-nature-13,christle-nature-14,singh-prb-20,Lekavicius-PRXQ-22}.
Debye-Waller factors of about 6\% were found for $V_{1}$ and $V_{2}$
type of $V_{Si}^{-}$$\;$\cite{Udvarhelyi-prappl-20,shang-prb-20},
which is vital for the quantum application that rely on inditinguishable
photon emitters. The electronically excited state zero-field splitting
of the $V_{2}$ centers in $4H$ -SiC exhibits a significant thermal
shift, which makes them useful for thermometry applications$\;$\cite{anisimov-sr-2018}.
$V_{Si}^{-}$ in SiC provides a low-cost and simple approach to quantum
sensing of magnetic fields and a sensitivity of 50 nT/$\sqrt{\text{Hz}}$,
which was achieved without complex photonic engineering, control protocols,
or applying excitation (optical and radio-frequency) powers greater
than a watt~\cite{anraham-prap-21}. Absolute dc magnetometry, which
is immune to thermal noise and strain inhomogeneity has been demonstrated
using all four ground-state spin levels of $V_{Si}^{-}$$\;$\cite{soltamov-naturecom-19}.
Similar to NV in diamond, $V_{Si}^{-}$ based magnetometers can also
be applied to unique magnetic sensing challenges, such as resolving
an individual nuclear spin in a molecule~\cite{taylor2008high}.
Since the magnetic field of a dipole decreases with distance as 1/$r^{3}$,
it is essential that the sensor can be brought close to the source
~\cite{taylor2008high}, which is only possible if the $V_{Si}^{-}$
is close to the surface.

Silicon vacancies can be created, e.g., by irradiating the sample
with neutrons or electrons~\cite{fuchs2015engineering,hahn-jap-14,kasper-prap-20}.
The charge on the silicon-vacancy depends on the Fermi-level: it is
at least 1.24 eV above the valence band, the $k$ lattice site becomes
negatively charged and at 1.5 eV also the $h$ lattice site$\;$\cite{hornos2011large}.
Irradiation by proton or other charged ions allow one to create a
plane of vacancies at a specific depth that depends upon the energy
and type of the ions$\;$\cite{babin2022fabrication,wang2017scalable,pavunny2021arrays}.
Here we focus on single silicon vacancy centers created with 6 keV
He$^{+}$ ion irradiation. Using a lithographic polymethyl methacrylate
(PMMA) mask with 100 nm holes, an array of silicon vacancies was created~\cite{babin2022fabrication}
at a depth of 30-40 nm below the surface. Additional details of the
sample preparation are given in Appendix A.

In this work, we focus on the $V_{Si}^{-}$ in the $4H$-SiC polytype,
specifically $V_{Si}^{-}$ centers at hexagonal sites $h$ and cubic
sites $k$, which are commonly known as $V_{1}$ and $V_{2}$$\;$\cite{viktor-prb-17}.
We characterize the properties of several individual centers and obtain
useful values for the optical transitions rates as well as for the
radiation-less population dynamics. We demonstrate that the optical
pumping results in spin polarization that is sufficiently high to
observe optically detected magnetic resonance (ODMR) spectra of single
silicon vacancy centers at room temperature with high contrast. From
a series of ODMR spectra with different RF powers, we could extract
the transverse dephasing time $T_{2}^{*}$.

This paper is organized as follows. Section \ref{sec:System} introduces
the optical and magnetic properties of a single spin system. Section~\ref{sec:Population-Dynamics}
provides details of correlation measurement experiments and analysis
to estimate the rates involved in the population dynamics of the center
during optical excitation. Section$\;$\ref{sec:Optically-Detected-Magnetic}
shows continuous-wave optically detected magnetic resonance (cw-ODMR)
experiments at different RF powers. Finally, Sec.\ref{sec:Discussion-and-Conclusion}
contains a brief discussion and some concluding remarks.

\section{System and Setup}

\label{sec:System}

\begin{figure}
\includegraphics[scale=1.1]{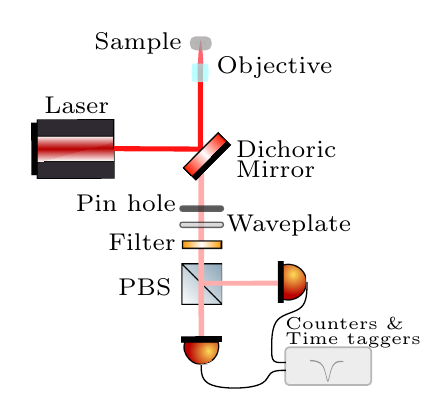}

\caption{(a) Experimental confocal setup used for PL and intensity autocorrelation
measurements.}

\label{pl-fig-1}
\end{figure}

The $V_{2}$ center has $C_{3v}$ symmetry and an electron spin $S=$
3/2. In the absence of a magnetic field, the spin Hamiltonian is

\begin{equation}
{\cal H}=D(S_{z}^{2}-\frac{S(S+1)}{3}),\label{eq:hamiltonian}
\end{equation}
where the zero field splitting (ZFS) in the electronic ground state
is $2D$= 70 MHz in frequency units ($h=1$)$\;$\cite{widmann-nature-14,baranov-prb-11}.
$\vec{S}$ = \{$S_{x},$ $S_{y}$, $S_{z}$\} is the vector of spin
operators and we have chosen the $z$-axis parallel to the $c$-axis
of the crystal ($C_{3}$ symmetry axis).

Figure$\;$\ref{pl-fig-1} shows the experimental confocal setup for
measuring photoluminescence (PL) and the second-order intensity autocorrelation.
A 785 nm laser beam was focused on the sample with an objective lens
of numerical aperture (NA) 1.3. With the help of a dichroic mirror,
the PL emitted by the sample was filtered and allowed to pass through
a $\lambda/2$ waveplate, a set of optical filters and a polarizing
beam splitter (PBS). Two single-photon detectors (SPD) were used for
the two possible paths. With the $\lambda/2$ waveplate, we controlled
the ratio of photons detected by these two detectors. From the arrival
time of the photons, we calculated the intensity autocorrelation function
$g^{(2)}$. For taking PL scans, the counts of both SPDs were added
and plotted against the XY position. For simplicity, some of the mirrors
and lenses are not shown in Fig.~\ref{odmr setup}, details are summarised
in appendix A.

Figure$\;$\ref{pl-fig} shows a confocal PL scan of a part of the
sample. The plot shows that most $V_{Si}^{-}$ centers lead to a photon
detection rate of $\sim$8000 cps (counts/s) at room temperature.
Higher count rates are also observed in some locations, indicating
multiple close centers. The counts due to impurities and dirt are
higher than the counts of the vacancies and usually bleach away after
a few minutes of laser illumination.

\begin{figure}
\includegraphics[scale=0.15]{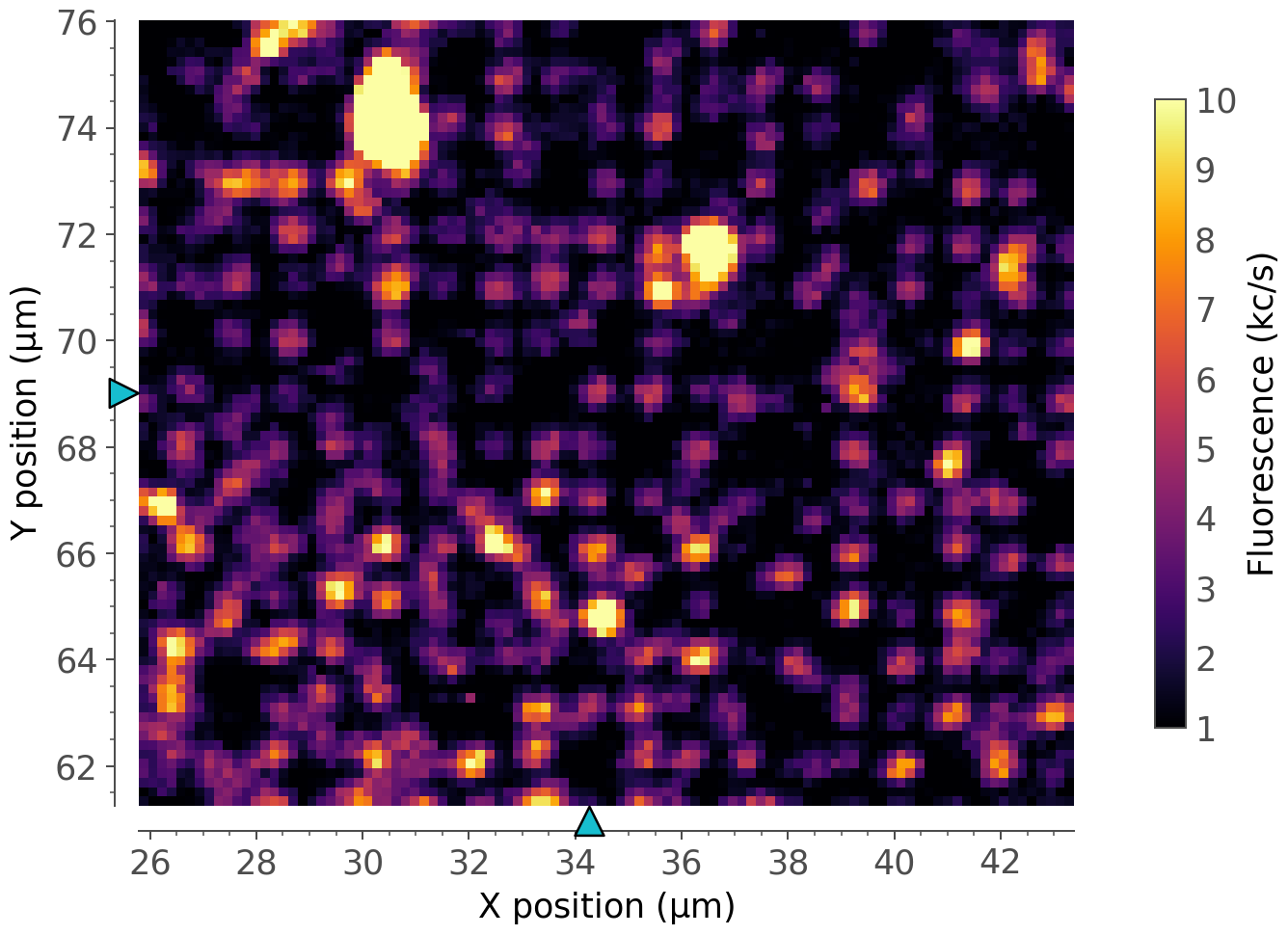}

\caption{Confocal scan of the defect center array in the XY plane. Large count
rates are due to dirt and impurities. The count rate is color-coded,
and its value can be taken from the bar on the right.}

\label{pl-fig}
\end{figure}

\section{Population Dynamics}

\label{sec:Population-Dynamics}

\begin{figure}
\includegraphics[scale=1.3]{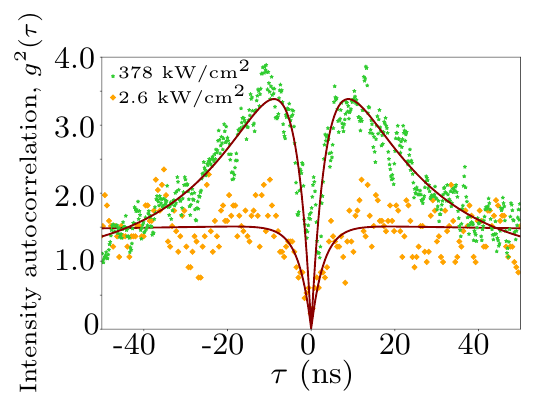}

\caption{Intensity autocorrelation measurements performed with 2.6 kW/cm$^{2}$
and 378 kW/cm$^{2}$ laser intensity in the focus. In the data recorded
with intensity $I$ = 2.6 kW/cm$^{2}$, the background count rate
was negligible, but it was significant for $I$=378 kW/cm$^{2}$.
We therefore performed a background correction. Each solid line follows
the three-level correlation function from Eq.$\;$\eqref{eq:g2}.}

\label{g2-exp}
\end{figure}

We measured the second-order correlation function of the PL light
emitted by a $V_{Si}^{-}$. This allowed us to verify that the light
is emitted by a single center as well as to study the population dynamics.
The second order correlation function
\begin{eqnarray*}
g^{(2)}(\tau) & = & \frac{\langle n_{1}(t)n_{2}(t+\tau)\rangle}{\langle n_{1}(t)\rangle\langle n_{2}(t+\tau)\rangle}
\end{eqnarray*}
is the conditional probability of measuring two photons with a delay
$\tau$. Here, $n_{i}(t)$ are the number of photons detected at time
$t$$\;$\cite{fox2006quantum}. For an ideal single photon emitter
the conditional probability of detecting two photons simultaneously
drops to zero, $g^{(2)}(\tau=0)$=0$\;$\cite{fox2006quantum}. Experimentally
$g^{(2)}(0)<0.5$ is evidence for a single photon emitter. Figures$\;$\ref{g2-exp}
and \ref{alg2} show the experimental correlation data. The minimum
in the curve is shifted from $\tau=0$ by $\tau_{0}=1.0\pm0.2$ ns,
which originates in unequal delays of the optical and electrical detection
paths of the two SPDs.

We performed correlation measurements with different laser intensities.
Figure$\;$\ref{g2-exp}$\,$, shows the intensity autocorrelation
vs. the delay $\tau$ for laser intensities 2.6 kW/cm$^{2}$ and 378
kW/ cm$^{2}$ with orange diamonds and green stars, respectively.
The bin width, count rates, and background details used for normalizing
the plots are given in appendix B along with data for additional laser
intensities. Experimental correlation functions with $I$\textless 135
kW/cm$^{2}$, reveal an antibunching dip, indicating that the selected
PL source is a single photon emitter $g^{(2)}(\tau=0)$ = 0.17 for
laser intensity $I$ = 2.6 kW/cm$^{2}$. As shown in Fig.~\ref{fig:Experimentally-obtained-PL-signa-1},
the signal increases linearly with the laser intensity $I$ for low
power but saturates for $I>$130 kW/cm$^{2}$. The background increases
linearly with $I$, which results in a decreasing signal-to-background
ratio at higher powers $I$ and a reduced antibunching dip ($g^{(2)}$
$(\tau=0)$ \textgreater{} 0.5), for data taken at $I$ \textgreater 135
kW/cm$^{2}$(plotted in Fig.~\ref{alg2} in appendix B)$\;$\cite{PhysRevLett.69.1516}.
A bunching behaviour is observed for $|\tau|$ \textgreater{} 9 ns
for the laser intensity of 378 kW/cm\texttwosuperior{} in Fig.$\;$\ref{g2-exp}).

\begin{figure}
\includegraphics{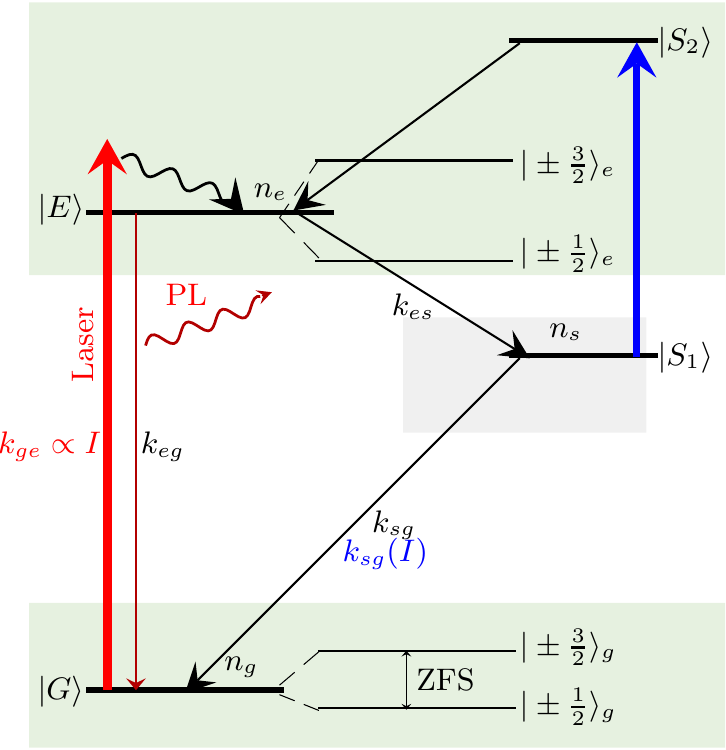}

\caption{Electronic energy levels scheme for $V_{Si}^{-}$ on the cubic site
($V_{2}$). The optical excitation with a 785 nm laser, from the ground
state to the excited state is indicated by the thick red arrow. Direct
radiative (PL) relaxation from the excited state $|E\rangle$ to the
ground state $\vert G\rangle$ is represented by a thin dark-red arrow.
From the excited state the center can undergo inter-system crossing
(ISC) to the shelving state $\vert S_{1}\rangle$, as indicated by
black arrows. The ground state spin-sublevels are separated by the
ZFS. $k_{ge}$, $k_{eg}$, $k_{es}$ and $k_{sg}$ are the transition
rates from the ground to the excited state, excited state to ground
state, excited state to shelving state and shelving state to ground
state, respectively. The deshelving process, corresponds to a transfer
of population from $\vert S_{1}\rangle$ to a higher energy level,
$\vert S_{2}\rangle$, followed by another ISC.}

\label{3-4level}
\end{figure}

These data can be modeled by considering at least a three-level system,
as shown in Fig.$\;$\ref{3-4level}. Absorption of laser light brings
the system from the ground state $\vert G\rangle$ to the excited
state $\vert E\rangle$ with a rate $k_{ge}$, which is proportional
to the laser intensity $I$. Due to spontaneous emission from the
excited state $\vert E\rangle$, most of the population returns to
the ground state $\vert G\rangle$ with a rate $k_{eg}$ and emits
PL. Some of the population goes to the shelving state $\vert S_{1}\rangle$
due to the intersystem crossing (ISC) with a rate $k_{es}$. From
the shelving state, the population returns to the ground state with
a rate $k_{sg}$. From the shelving state population preferentially
populates the $\vert\pm1/2\rangle$ spin substates. The resulting
population difference of up to 80\%~\cite{soltamov-prl-12} makes
ODMR measurements possible. Writing $n_{g}$, $n_{e}$, and $n_{s}$
for the populations of the electronic ground state $\vert G\rangle$,
excited state $\vert E\rangle$, and shelving state $\vert S_{1}\rangle$
respectively, the population dynamics can be written as
\begin{eqnarray}
\frac{d}{dt}\vec{n}=\frac{d}{dt}\left(\begin{array}{c}
n_{g}\\
n_{e}\\
n_{s}
\end{array}\right)= & \left(\begin{array}{ccc}
-k_{ge} & k_{eg} & k_{sg}\\
k_{ge} & -k_{eg}-k_{es}\\
 & k_{es} & -k_{sg}
\end{array}\right) & \vec{n}.\label{eq:three-levelrate}
\end{eqnarray}

The eigenvectors $\vec{v_{i}}$ are given in appendix D and the eigenvalues
for Eq.\eqref{eq:three-levelrate} are

\begin{eqnarray}
\vec{\lambda} & =\frac{1}{2} & \left(\begin{array}{c}
0\\
A+\sqrt{A^{2}-4B}\\
A-\sqrt{A^{2}-4B}
\end{array}\right),\label{eq:egenvalues}
\end{eqnarray}

where $A$ = $k_{ge}+k_{eg}+k_{es}+k_{sg}$ and $B$ = $k_{sg}(k_{eg}+k_{es}+k_{ge})+k_{es}k_{ge}$.
The resulting stationary state is

\begin{eqnarray}
\vec{n}^{st} & = & \frac{1}{B}\left(\begin{array}{c}
k_{sg}(k_{eg}+k_{es})\\
k_{ge}k_{sg}\\
k_{es}k_{ge}
\end{array}\right).\label{eq:stationary}
\end{eqnarray}

To calculate the theoretical correlation function $g^{2}(\tau)$,
we need the excited state population at time $\tau$ when the system
is initially in the ground state ($n_{g}(0)=1$ and $n_{e}(0)=n_{s}(0)=0$):
\begin{eqnarray}
\frac{n_{e}(t)}{n_{e}^{st}} & \overset{t=\tau}{=}g^{(2)}(\tau)= & 1-(1+c)e^{-\ensuremath{\tau}/\ensuremath{\tau_{1}}}+c\;e^{-\ensuremath{\tau}/\ensuremath{\tau_{2}}},\label{eq:g2}
\end{eqnarray}
with the inverse of the eigenvalues 
\begin{equation}
\tau_{1,2}=2/(A\pm\sqrt{A^{2}-4B})\label{eq:tau_eigen}
\end{equation}
 and the bunching amplitude
\begin{eqnarray}
c & = & \frac{1-\tau_{2}k_{sg}}{k_{sg}(\tau_{2}-\tau_{1})}.\label{eq:c}
\end{eqnarray}

\begin{figure}
\includegraphics{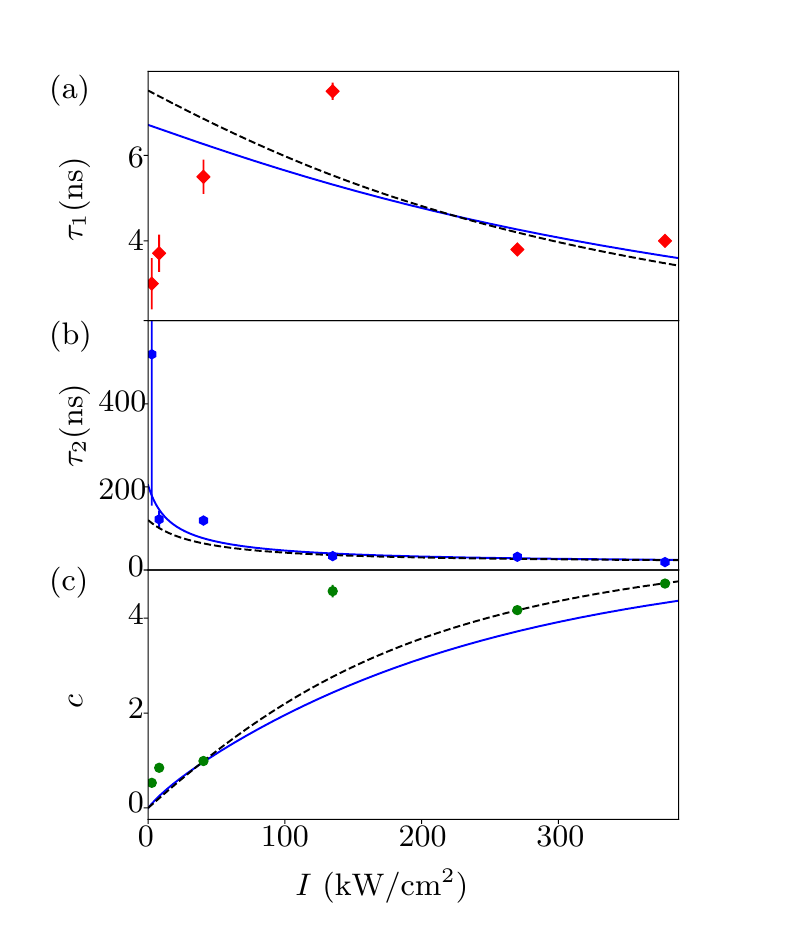}

\caption{Time constants $\tau_{1}$, $\tau_{2}$ and bunching amplitude $c$
of the second order correlation function for various laser power intensities.
The experimental data are fitted with the functions of Eqs.~\eqref{eq:egenvalues}
and \eqref{eq:c}. All three datasets are fitted simultaneously which
results in the following parameters $\tau_{1}^{0}$ = (7.5$\pm$ 1.1)
ns, $\tau_{2}^{\infty}$ =(17.2 $\pm$1.4) ns and $c^{\infty}$ =
6.0 $\pm$ 0.5 for the 3-level model (black dashed curves); $\tau_{1}^{0}$
= (6.7 $\pm$ 1.4) ns, $\tau_{2}^{0}$ =(204.4 $\pm$ 166.3) ns, $\tau_{2}^{\infty}$
=(14.9 $\pm$ 3.4) ns and $c^{\infty}$ = 6.3 $\pm$ 0.9 for the 4-level
model (blue curves).}

\label{param}
\end{figure}

In Fig.~\ref{g2-exp}, the solid dark red lines are the fitted curves
obtained using Eq.$\;$\eqref{eq:g2}. In Fig.\,\ref{param}, we
plot the extracted fit parameters $\tau_{1}$, $\tau_{2}$, and $c$
vs. the laser intensity. The experimental data points are compared
to theoretical curves that can be obtained from the intensity dependence
of the effective rate coefficients

\begin{eqnarray}
k_{sg} & = & \frac{1}{(1+c^{\infty})\tau_{2}^{\infty}},\nonumber \\
k_{es} & = & k_{sg}c^{\infty},\nonumber \\
k_{eg} & = & \frac{1}{\tau_{1}^{0}}-k_{23}\,\,(k_{eg}+k_{es})>k_{sg},\label{eq:inlow}
\end{eqnarray}
where the values with superscript $0$ refer to the low-intensity
limit and those with superscript $\infty$ to the high intensity limit~\cite{Neu_2011,fuchs2015engineering}.
The absorption cross section $\sigma$ was obtained from the transition
rate $k_{ge}$ from the ground state to the excited state:

\begin{eqnarray*}
\sigma & = & \frac{h\nu}{I}(\frac{k_{ge}}{n_{g}})
\end{eqnarray*}
with the population $n_{g}$ of the ground state and $I/h\nu$ the
number of incident photons per second per unit area. At saturation
intensity $I_{0}$, the population $n_{g}$ drops to 1/2:

\begin{equation}
\frac{k_{ge}}{n_{g}(\infty)}=\frac{k_{sg}\left(k_{eg}+k_{es}\right)}{k_{sg}+k_{eg}}.
\end{equation}

The absorption cross section can be written as

\[
\sigma=\frac{k_{ge}}{n_{g}}\frac{h\text{\ensuremath{\nu}}}{I_{0}}=\frac{k_{sg}\left(k_{eg}+k_{es}\right)}{k_{sg}+k_{es}}\frac{h\text{\ensuremath{\nu}}}{I_{0}},
\]

Fitting simultaneously the obtained values of $\tau_{1}^{0}$ = (7.5$\pm$
1.1) ns, $\tau_{2}^{\infty}$ =(17.2 $\pm$1.4) ns and $c^{\infty}$
= 6.0 $\pm$ 0.5 from the data plotted in Fig.~\ref{param} after
simultaneously fitting in Eqs.~\eqref{eq:egenvalues} and \eqref{eq:c}.
As in previous works~\cite{fuchs2015engineering,Neu_2011}, the three-level
model describes well the antibunching decay time constant $\tau_{1}$
(the black dashed line in Fig.~\ref{param}(a)) and the bunching
amplitude $c\,(I)$ (the black dashed line in Fig.~\ref{param}~(c)).
Table~\ref{rates} shows the calculated rates and absorption cross
section obtained from Eqs.~\eqref{eq:inlow}. For low intensity ($I$
\textless{} 104 kW/ cm\textsuperscript{2}) the bunching time constant
reaches several hundred ns, much longer than the window considered
in Fig.~\ref{g2-exp}. At higher intensities, the bunching time gets
shorter. In Ref. \cite{Neu_2011,fuchs2015engineering}, this is explained
as a de-shelving process that can be described as shown in Fig.~\ref{3-4level}:
The laser re-excites the system from the shelving state $\vert S_{1}\rangle$
to a higher-lying state $\vert S_{2}\rangle$ from where it can fall
back to $|E\rangle$. The following are the rates for the 4-level
model

\begin{eqnarray}
k_{sg} & = & \frac{d\,I}{I+I_{0}}+k_{sg}^{0},\nonumber \\
k_{sg}^{0} & = & \frac{1}{\tau_{2}^{0}},\nonumber \\
k_{sg}^{\infty} & = & \frac{1}{\tau_{2}^{\infty}},\nonumber \\
k_{es} & = & \frac{1}{\tau_{2}^{\infty}}-k_{sg}^{0}-d\nonumber \\
k_{eg} & = & \frac{1}{\tau_{1}^{0}}-k_{es},\label{eq:4levels}
\end{eqnarray}
where $d=\frac{1/\tau_{2}^{\infty}-(1+c^{\infty})/\tau_{2}^{0}}{1+c^{\infty}}$~\cite{Neu_2011,fuchs2015engineering}.
Inclusion of the deshelving process results in the intensity-dependent
rate from shelving state to ground state $k_{sg}$~\cite{Neu_2011}.
Fitting simultaneously the obtained values of $\tau_{1}^{0}$ = (6.7
$\pm$ 1.4) ns, $\tau_{2}^{0}$ =(204.4 $\pm$ 166.3) ns, $\tau_{2}^{\infty}$
=(14.9 $\pm$ 3.4) ns and $c^{\infty}$ = 6.3 $\pm$ 0.9 from the
data plotted in Fig.~\eqref{param} after simultaneously fitting
in Eqs.~\eqref{eq:egenvalues} and \eqref{eq:c}, using 4-level rates
given in Eqs.~\eqref{eq:4levels}. The blue curves in Fig.~\ref{param}
correspond to the 4-level model, which matches the experimental data
better than the black dashed curves from the 3-level model. We now
compare our results to previous work~\cite{fuchs2015engineering},
in which silicon vacancies were created through high-energy neutron
irradiation without post-annealing. The extracted transition times
and cross-sections from the 4-level model were $1/k_{eg}$=7.6 ns,
1/$k_{es}$=16.8 ns, 1/$k_{sg}^{0}$=150 ns, 1/$k_{sg}^{\infty}$=123
ns, and $\sigma$= 1.6 $\times$ 10$^{-16}$cm\textsuperscript{2}~\cite{fuchs2015engineering}.
The transition times $1/k_{eg}$, 1/$k_{es}$ and $1/k_{sg}^{0}$
measured by us for the shallow silicon vacancy center are compatible,
within the experimental uncertainties, with those of Fuchs \emph{et
al}.~\cite{fuchs2015engineering}. Further the cross-section $\sigma$
error bar's lower end is close to the cross-section measured by Fuchs
et al.~\cite{fuchs2015engineering}. The main difference between
the two samples is the defect generation method and the depth of the
center. We tentatively assign the defect generation method to the
$\approx5\times$ increase that we observed in our rate $k_{sg}^{\infty}$.
In our sample, emitters are created with low-energy Helium ion implantation,
followed by 600°C annealing to remove parasitic lattice damage. As
has been shown in previous work~\cite{babin2022fabrication}, high
damage defect generation can significantly alter radiative and non-radiative
decay processes, which may modify relaxation rates~\cite{kasper2021engineering}.
Since our emitters are located within $\approx$40 nm at the surface,
we additionally expect coupling to surface charge states and traps,
which may lead to a symmetry reduction through which additional decay
channels can be allowed. We believe that a more thorough analysis
on the modification of the rates as a function of the defect generation
method would be useful, however, this would require low-temperature
experiments~\cite{PhysRevApplied.17.054005}, which go beyond the
scope of this paper.

Over the duration of the intensity autocorrelation measurement, the
count rate of the selected single center dropped significantly. We
therefore used a different center (we call it center 2) for the ODMR
experiment. The intensity autocorrelation measurement data for that
center confirm that it is also a single center.

\begin{table}
\begin{centering}
\begin{tabular}{|c|c|c|c|c|c|}
\hline 
\multicolumn{5}{|c|}{Inverse rate coefficients in ns} & Cross section\tabularnewline
\hline 
Model & $1/k_{eg}$ & $1/k_{es}$ & $1/k_{sg}^{0}$ & $1/k_{sg}^{\infty}$ & $\sigma$ ($10^{-16}$cm$^{2}$)\tabularnewline
\hline 
\hline 
3 levels & 12 $\pm$ 3 & 20 $\pm$ 2 & - & 120 $\pm$ 13 & 1.1$\pm$.2\tabularnewline
\hline 
4 levels & 11 $\pm$ 4 & 17 $\pm$ 8 & 204 $\pm$ 166 & 15 $\pm$ 3 & 4.9 $\pm$ 1.2\tabularnewline
\hline 
\end{tabular}
\par\end{centering}
\caption{Inverse rate coefficients and absorption cross section of the $V_{Si}^{-}$
center.}

\label{rates}
\end{table}

\section{Optically Detected Magnetic Resonance}

\label{sec:Optically-Detected-Magnetic}

In conventional electron spin resonance (ESR), spins are measured
by inductive detection, which limits the sensitivity of the technique.
Using optical polarization of the spin system and optical detection,
it is possible to sufficiently increase the sensitivity for experiments
on single spins \cite{mr-1-115-2020,SUTER201750}. In this optically
detected magnetic resonance (ODMR), the electron spin is excited with
laser light to its optical excited state, and the PL emitted during
re-emission is detected while a radio-frequency (RF) or microwave
(MW) field drives the spin system. The different spin states contribute
differently to the PL rate, so that a change in the spin polarization
leads to a change of the PL rate, which can be used to measure spin
polarization. We used this technique for measuring the ODMR of a single
silicon vacancy in $4H$-SiC.

For recording the single center's continuous wave (CW) ODMR, the center
was continuously illuminated with an optimized laser intensity of
68 kW/cm$^{2}$ where the signal to background ratio is maximal; more
details are given in Appendix F. The ODMR was recorded in ``lock-in
mode'': while stepping the RF over the relevant frequency range,
PL counts were recorded with and without RF. The difference between
the two vlaues \textgreek{D}PL = PL$_{RF}$\textminus PL$_{off}$
, is less sensitive to fluctuations of the laser intensity.

\begin{figure}
\includegraphics[scale=1.2]{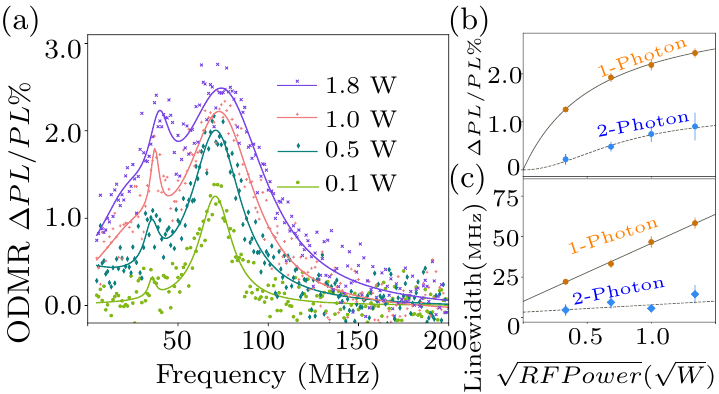}

\caption{(a) ODMR signal vs. frequency recorded for center 2 with different
RF powers in absence of applied magnetic fields. The horizontal axis
is the frequency in MHz, and the vertical axis is the relative change
of the PL. (b) ODMR signal vs. RF field strength and (c) linewidth
vs. RF field strength.}

\label{odmr}
\end{figure}

Figure~\ref{odmr}~(a) shows the recorded ODMR signal at different
RF powers (0.1 W, 0.5 W, 1.0 W and 1.8 W). The ODMR spectrum recorded
with 0.1 W fitted in double Lorentzian function, and the obtained
resonance frequencies are (72.8 $\pm$ 0.3) MHz for the main peak
and (36.4 $\pm$ 0.2) MHz for the smaller peak.The frequency of the
larger peak matches the literature value of the zero-field splitting
of the $V_{2}$ type $V_{Si}^{-}$ in $4H$-SiC~\cite{baranov-prb-11}.
At high RF power the Rabi frequency of the system is comparable to
the resonance frequency of the system, the rotating wave approximation
is no longer valid and causes nonlinear processes, such as multi-RF
photon absorption~\cite{singh-prr-2022}. In our previous work~\cite{singh-prr-2022},
we studied the multi-RF process in detail. In those experiment, the
large signal-to-noise ratio of the $V_{3}$ spin ensemble in allowed
detailed studies of the 1-, 2-, 3-, and 4-RF photon resonances. Here,
the smaller peak is at half the frequency of the main peak, so we
attribute it to the 2-RF photon peak. In the simulated spectra shown
in Appendix G, Fig~\ref{sim_odmr} (a), the multi-photon peaks are
more clearly visible. In the experiment, the signal due to higher
multi-photon absorption are not resolved due to limited signal-to-noise
ratio. To take these unresolved signal contributions into account,
we add a third Lorentzian component to the fitting function of the
ODMR spectra when the applied RF power is higher than 0.5 W. Figures~\ref{odmr}~(b)
and (c) show the RF power dependence of the amplitudes and the linewidths
for 1- and 2-photons peaks. The dependence of the amplitude is fitted
with the function

\begin{equation}
S(P)=S_{max}[(P^{c/2}/(\varLambda_{0}+P^{c/2})],\label{eq:signal}
\end{equation}
where S(P) is the ODMR signal amplitude and $P$ is the RF power.
The exponent $c$ is 1 (2) for the 1- (2)-photon transitions~\cite{singh-prr-2022}.
The values for $S_{max}$ and $\varLambda_{0}$ for the peaks are
given in Table~\ref{fittingparameter}.

The linewidth $LW$ vs. the square root of the RF power $\sqrt{P}$
is fitted to the function

\begin{equation}
LW(P)=LW_{0}+a\,\sqrt{P}.\label{eq:linewidth}
\end{equation}
 The obtained fitting parameters, intercept $LW_{0}$ and slope $a$
are given in Table~\ref{fittingparameter}. The low-power limit $LW_{0}$
corresponds to the inverse dephasing time $1/(\pi LW_{0})$~\cite{wertz2012electron}.
The low-power limit of the linewidth corresponds to a dephasing time
of $32\pm7$ ns. One contribution to the linewidth are stray magnetic
fields of about (0.5 G), which cause extra 2.8 MHz peak broadening.
Apart from that, the contrast and the linewidth depend on the RF and
laser intensity~\cite{jensen-prb-13}. For the same kind of centers,
$T_{2}^{*}$ = 34 $\pm$ 4 $\mu$s was measured at 10 K using Ramsey
interferometry in an external magnetic field of $B_{0}$ = 36 G oriented
along the crystal\textquoteright s $c$-axis~\cite{babin2022fabrication},
and similar values were obtained in deep-bulk defects in similar SiC
crystals~\cite{bourassa2020entanglement,nagy2019high}. Due to differences
in experimental conditions between the present work and the study
by Babin et al.~\cite{babin2022fabrication}, a quantitative comparison
of the dephasing rates estimated here from the low-power limit, $LW_{0}$,
and the $T_{2}^{*}$ measured in their study is not meaningful. However,
it is worth considering some conditions contributing to the differences.
Babin et al. performed their measurements at a low temperature of
10 K, while the measurements presented here were obtained at room
temperature. Additionally, Babin et al. used the Ramsey method to
measure free precession, whereas the present measurements were obtained
using the CW method. Furthermore, in Babin et al.'s study, selective
low-power resonant excitation ensured the system was predominantly
in the ground state. In contrast, in the present experiment, CW excitation
led to the system being mostly in the metastable state, which contributes
to the dephasing and line broadening. Lastly, Babin et al. applied
a magnetic field to lift the degeneracy of the electron spin states,
which reduces line-broadening from magnetic field noise and changes
the dynamics of the nuclear spin bath and its contribution to the
line broadening~\cite{bulancea2021dipolar}. Fig.~\ref{sim_odmr}~(b)
of Appendix H shows the linewidth of the simulated spectra vs. the
RF coupling strength ($\Omega_{1}=g\mu_{B}B_{1})$, which looks similar
to the experimental data in Fig.~\ref{odmr}~(c). As shown in Fig~\ref{sim_odmr}~(b)
the width of the 2-photon resonance for low power tends to zero (which
corresponds to $T_{2}^{*}$ $\longrightarrow$ $\infty$ for 2-photon
resonance) and the slope $a_{1}$ of the 2-photon resonance is smaller
than that of the 1-photon resonance.

\begin{table}
\begin{centering}
\begin{tabular}{|c|c|c|c|c|c|}
\hline 
Peak & $c$ & $S_{max}$(\%) & $\varLambda_{0}$($W^{c/2}$) & $LW_{0}$ (MHz) & $a$ (MHz/$\sqrt{W}$)\tabularnewline
\hline 
\hline 
$1$-photon & 1 & 3.51$\pm$0.25 & 0.59$\pm$0.10 & 10.0 $\pm$ 2.3 & 35.9$\pm$3.3\tabularnewline
\hline 
$2$-photon & 2 & 1.2$\pm$0.6 & 0.6$\pm$0.6 & 3.3$_{+4.0}^{-3.3}$ & 5$\pm$5\tabularnewline
\hline 
\end{tabular}
\par\end{centering}
\caption{Obtained fitting parameters from Eq.~\eqref{eq:signal} for the ODMR-signal
amplitude and Eq.~\eqref{eq:linewidth} for the linewidth vs RF field
strength.}

\label{fittingparameter}
\end{table}

\begin{figure}
\includegraphics{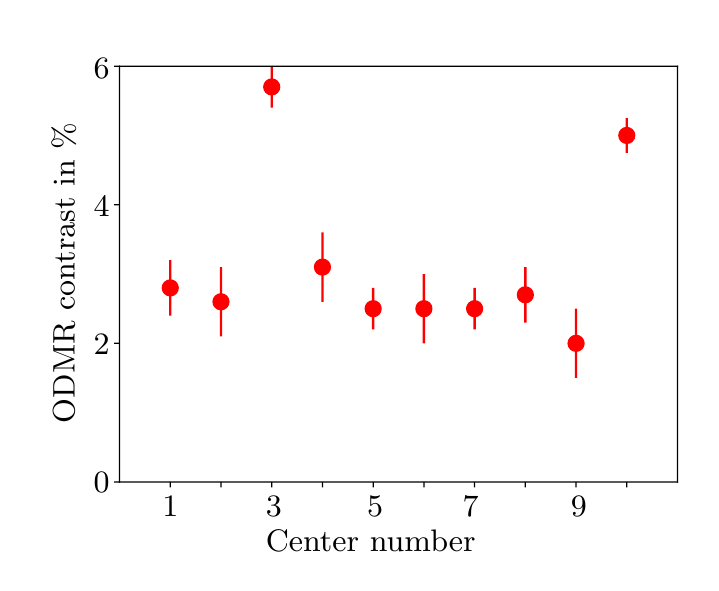}

\caption{ODMR contrast for different centers.}

\label{odmrcontrast}
\end{figure}

We recorded the ODMR of ten more centers with 9 W RF power and 185
$k$W/cm$^{2}$ laser intensity. We observed the contrast range and
Fig.~\ref{odmrcontrast} shows the plot of number of centers vs.
ODMR contrast. On average, we observed a $\approx$ 3 \% contrast
per center, with a minimum contrast of 1\% and a maximum contrast
of 6\%. We also recorded ODMR spectra of one of the high contrast
center for the different RF and laser Powers; more details are given
in appendix G.

\section{Discussion and Conclusion}

\label{sec:Discussion-and-Conclusion}

Vacancy centers in silicon carbide are useful tools for sensing magnetic
fields and for other quantum technologies. With existing fabrication
technology, $V_{Si}^{-}$ can be produced on-demand with a high probability,
and reasonable resolution. Previously, the spin properties of a $V_{2}$
center deep inside a $4H$-SiC sample have been studied using a solid
immersion lens to enhance the PL of the $V_{2}$ center$\;$\cite{widmann-nature-14}.
Here we characterized shallow single silicon-vacancy centers, which
are very useful for quantum sensing applications. The single center
is located approximately 40 nm below the surface, can be optically
spin-polarized, and the measured optical transition rates are slightly
different from the previously measured deep silicon-vacancy centers$\;$\cite{widmann-nature-14}
and fitted well in 4-level model$\;$\cite{fuchs2015engineering,Neu_2011}.

The $T_{2}^{*}$ of transition $\pm$3/2 $\longleftrightarrow$$\pm$1/2
was 32 ns. Longer dephasing times are expected for the +1/2 $\longleftrightarrow$
-1/2 transition, since it is not affected by the zero-field splitting~\cite{soltamov-naturecom-19,singh-prb-21};
this may be interesting for sensing applications. Observing this transition
is possible in an external magnetic field, which lifts the degeneracy
of these levels and can result in a population imbalance between them.
The maximum ODMR contrast of $\approx$6 \% was obtained with 9 W
RF power and 185 $k$W/cm$^{2}$ laser intensity.

To understand the exact cause of the difference between these transition
rates and the ODMR contrast of the shallow and deep centers in our
future work, we plan to perform experiments where we change the depth
of the center, e.g. by etching techique. Illumination for more than
24 hours with an intensity of 378 kW/ cm\texttwosuperior{} reduces
the count rates. The reduced count rate may be due to charge state
conversion~\cite{wolfowicz2017optical}. Here we used a 785 nm laser,
i.e. non-resonant pumping. It may be possible to avoid the reduction
of the count rate by using a different laser wavelength close to zero
phonon line. We will try this approach in our future work.

In conclusion, we reported room temperature ODMR data from single
shallow $V_{Si}^{-}$ centers in $4H$-SiC. We observed higher ODMR
contrast from in the centers that were created with He\textsuperscript{+}
implantation than in previous studies where centers had been created
by neutron irradiation~\cite{widmann-nature-14,wang2019demand,fuchs2015engineering}.
Due to the higher contrast, we are able to record ODMR of a center
with PL $\approx$ 8$k$cps. An ensemble of $V_{si}^{-}$ in SiC has
a single dipole orientation, in contrast to an ensemble of NV centers
in diamond with four possible dipole orientations. For some applications,
this increases the signal by a factor of four. So, this increased
ODMR contrastdue to shallowness makes the silicon-vacancy center
a good candidate as NV center in diamonds for room temperatures sensing
applications such as NMR of a small sample volume and MRI of single
cells~\cite{glenn2015single,taylor2008high,glenn2018high,PhysRevX.10.021053}.

\section*{Appendix A: Sample}

Arrays of $V_{Si}^{-}$ centres via implantation of He$^{+}$ ions.
He$^{+}$ ions were sent through a polymethyl methacryate (PMMA) mask
with 100 nm diameter holes, lithographed on the a face of an epitaxially
grown SiC sample with a low nitrogen concentration in the {[}N{]}
= 4 $\times$ 10$^{13}$ cm$^{-3}$ range. The low He$^{+}$ ion energy
of 6 keV was chosen to minimize the crystal damage. To remove residual
lattice damage, the sample was annealed in argon atmosphere at 600°
C for 30 min~\cite{babin2022fabrication}. Figure$\;\ref{pl-fig}$
shows the room-temperature PL map of an implanted array using off-resonant
excitation at 785 nm.

\section*{Appendix B: ODMR setup}

\begin{figure}
\includegraphics[scale=0.5]{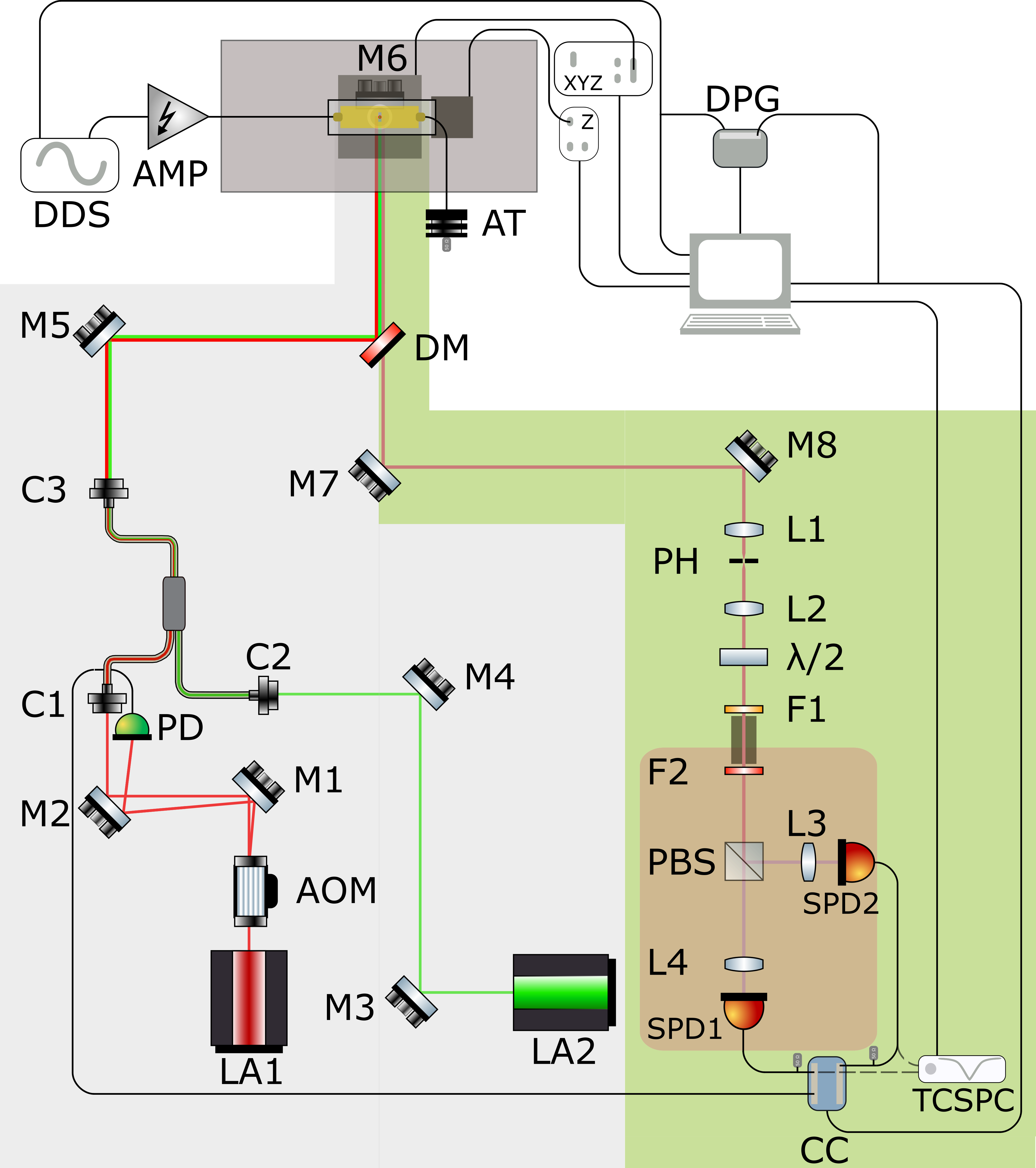}

\caption{Experimental confocal scan setup for the CW-ODMR (zfODMR), intensity
autocorrelation and pulse measurements. The detectors are isolated
in a box (in reddish-brown) to prevent noise from external light sources.
PL can enter the box via a tube (between F1 and F2). Excitation path
(gray background): the center is optically excited by a 785 nm laser.
RF transitions are induced by a 50 $\mu$m copper wire over the sample.
Detection path (green background): the PL is detected by two single-photon
detectors (SPD1 and SPD2), which are connected to the DAQ-card (DAQ)
for ODMR measurements. When performing a $g^{2}$-measurement, the
detectors are connected to the quPSI TCSPC device (indicated by dashed
lines).}

\label{odmr setup}
\end{figure}

The excitation path (Fig.$\;$\ref{odmr setup}, grey background)
begins with a 785 nm 250 mW iBEAM-SMART-785-S-HP Laser (LA1) from
Toptica which induces electronic transitions of the $4H$-SiC-system.
The 520 nm Lasertack 100 mW single mode diode laser (LA2) is used
for alignment. Each laser beam is directed by two mirrors (M1 and
M2 / M3 and M4) and coupled by the Thorlabs fiberports (C1/C2) into
a Thorlabs 532 nm/ 785 nm NG71F1 y-fiber which combines both beams.
To perform pulse experiments, an Acousto-Optic-Modulator (AOM) is
placed in front of the red laser (LA1). The zero order beam is detected
by a S2386-44K6K photodiode from Hamamatsu which monitors the stability
of the laser output. The 1$^{st}$ order diffracted beam is coupled
into the y-fiber. Finally the combined laser beam is coupled out of
the y-fiber by fiberport (C3) and directed by mirror (M5) to the DMLP805
long-pass dichroic mirror (DM) from Thorlabs, which transmits wavelengths
greater than 805 nm. Hence the combined light beam is reflected and
directed by mirror (M6) into an Olympus UPLFLN100XO2 objective. It
is mounted on a PI P-733.3CD XYZ-nanopositioner which covers a range
of 100 $\mu$m in X-/ Y- and 10 $\mu$m in Z-direction. This XYZ-nanopositioner
is operated by a PI E-727 Digital Multi-channel Piezo Controller.
The combined laser beam is focused onto the sample which is mounted
onto a circuit board, where RF-fields are applied through a 50 $\mu$m
copper wire. These RF-fields are generated by an Analog Devices AD9915
2.5 GSPS Direct Digital Synthesiser (DDS). The RF-power of range 20
MHz \textminus{} 512 MHz can be increased for several watts by a 50
W Mini Circuit LZY-1 amplifier (AMP). It is important to connect a
50 $\Omega$ resistor at the other end of the circuit board to prevent
damage of the electronic devices due to RF-reflections. Since the
resistor is limited to 1 W, the RF-power is increased by an attenuator
(AT) beforehand. The circuit board itself is screwed tight onto the
sample stage that allows movements in Z-direction by a motorised Standa
8MT173 translation stage which is operated by the Standa 8SMC5-USB-B8-1
motor controller. In Fig. $\;$\ref{odmr setup}, the circuit board
is shown in yellow whereas the objective is illustrated as a circle
below it. All components of the detection path (Fig.$\;$\ref{odmr setup},
green background) are selected for the near-infrared range of 805
nm-1000 nm. PL emitted from the sample is collected by the microscope
objective and directed by the mirror (M6) through the dichroic mirror
(DM) (bright red line in Fig.$\;$\ref{odmr setup}). It is then directed
by the mirrors (M7) and (M8) and focused by the lens (L1) with focal
length of 100 mm, passes a 50 \textgreek{m}m pinhole (PH) and is then
collimated by the lens (L2) with a focal length of 150 mm. The following
waveplate ($\lambda$/2) allows to adjust the ratio between transmitted
and reflected PL at the polarized beam splitter (PBS) in the box.
Before entering the box via a tube, the PL is filtered by a 850 nm
long-pass-filter (F1). Depending on the type of measurement, a second
filter (F2) like a (950 $\pm$ 25) nm band-pass-filter or a 850 nm
long-pass-filter is used. The PL is splitted by the PBS and directed
to the lens (L3/ L4), each with focal length of 100 mm, which focuses
the PL onto the Laser Components Count-100N single photon detectors
(SPD1 and SPD2, with photon detection efficiency at 810 nm is 68 \%
and 59\%). When performing an ODMR-measurement, the registered events
are counted by a Measurement Computing 1808X-USB DAQ-card (DAQ), which
has a 50 $\Omega$ resistor at each input to prevent reflections.
For intensity autocorrelation measurements, we used time tagger and
counters of quPSI from qu\textgreek{t}ools.

\section*{Appendix C: Details of intensity autocorrelation measurements}

\begin{figure}
\includegraphics[scale=1.5]{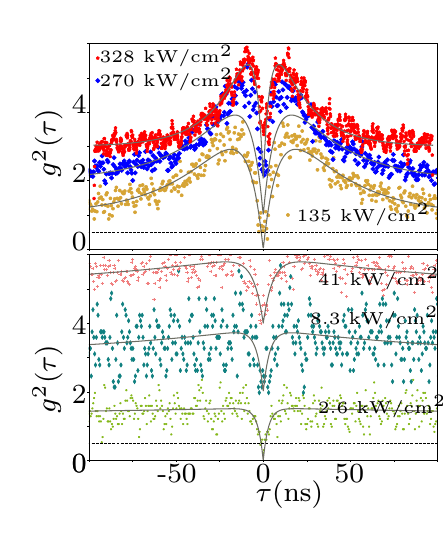}

\caption{Complete series of intensity autocorrelation measurements performed
from 2.6 kW/cm$^{2}$ to 378 kW/cm$^{2}$. The histogram data is first
normalized by $N$\protect\textsubscript{norm} = T $W_{b}$ $R_{1}$
$R_{2}$ and then smoothed to a bin width of 972 ps. Besides for the
lowest $I$ = 2.6 kW/cm\protect\textsuperscript{2}, all histogram
dataset are corrected for their background. Each solid line follows
the three-level correlation function from Eq.~\eqref{eq:g2}.}

\label{alg2}
\end{figure}

\begin{figure}
\includegraphics[scale=1.2]{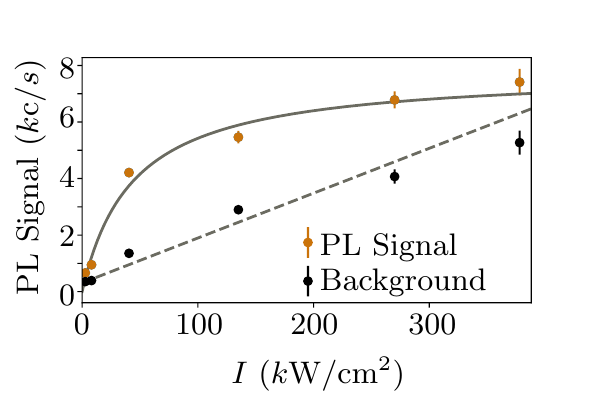}

\caption{PL-signal at different intensities $I$ of the center on which $g^{2}$-measurements
were performed. The PL-signal is fitted by: $S=S_{max}I/(I_{0}+I)$
with (7.8 $\pm$ 0.3) kcounts/s and $I_{0}$= (44 $\pm$ 3) kW/cm$^{2}$,
whereas the background increases linearly with the slope m = (15.9
$\pm$ 0.6) counts cm$^{2}$/(kW s) and intercept b = (309 $\pm$
20) counts/s.\label{fig:Experimentally-obtained-PL-signa-1}}
\end{figure}

All but one $g^{2}$-measurements shown in Fig.~\ref{alg2}  were
taken with a bin width $W_{b}$ of 486 ps, whereby the histogram data
for $I$ = 378 kW/cm$^{2}$ was obtained with W of 162 ps. Eventually,
all unnormalized histogram datasets were normalized by $N_{norm}$
= $T$ $W_{b}$ $R_{1}$$R_{2}$ , with total time $T$, bin width
$W_{b}$ and count rates $R_{i}$ of the detectors $D_{i}$ and afterwards
smoothed to a bin width of 972 ps. The total count rates $R$ = $R_{1}$
+ $R_{2}$ are obtained by dividing the total number of counts by
the total time for the $g^{2}$-measurement. Finally, the individual
count rates were then obtained by $R_{1}$ = 0.6~$R$ and $R_{2}$
= 0.4 $R$. At $I$ = 2.6 kW/cm$^{2}$, only the dark count rates
of $R_{1}$, D = 251 counts/s and $R_{2}$, $D$ = 129 counts/s at
each detector contribute to the background due to slightly different
models. But a linear increase in background for increasing excitation
power is observed in Fig.$\:$\ref{fig:Experimentally-obtained-PL-signa-1},
which can be considered by applying a background-correction. Since
the background-correction cannot be applied for a background which
only consists of the detectors dark counts, it was applied on all
histogram datasets except for the one at lowest excitation powers
of 2.6 kW/cm$^{2}$. Only the obtained normalized histogram data at
smallest (orange diamond) and at greatest (green star) optical excitation
power are shown in Fig.$\:$\ref{g2-exp}. All histogram datasets
are shown in Fig.$\:\ref{alg2}$.

\section*{Appendix D: Eigenvectors}

Eigenvectors of the three level model

$\vec{v}_{1}=\left(\begin{array}{c}
1\\
\frac{k_{sg}}{k_{es}}\\
\frac{k_{sg}(k_{eg}+k_{es})}{k_{es}k_{ge}}
\end{array}\right)$ , $\vec{v}_{2}=\left(\begin{array}{c}
-1\\
G\\
-F
\end{array}\right)$, $\vec{v}_{3}=\left(\begin{array}{c}
-1\\
-G\\
F
\end{array}\right)$,

where

$F=\frac{H+k_{eg}-k_{es}+k_{ge}-k_{sg}}{2k_{es}}$;

$G=\frac{H+k_{eg}+k_{es}+k_{ge}-k_{sg}}{2k_{es}}$

$H=\sqrt{k_{eg}^{2}+2k_{eg}(k_{es}+k_{ge}-k_{sg})-J}$

$J=2\,k_{sg}(k_{es}+k_{ge})+(k_{es}-k_{ge})^{2}+k_{sg}^{2}$

\section*{Appendix F: Optimal laser power for ODMR}

\begin{figure}
\includegraphics[scale=1.2]{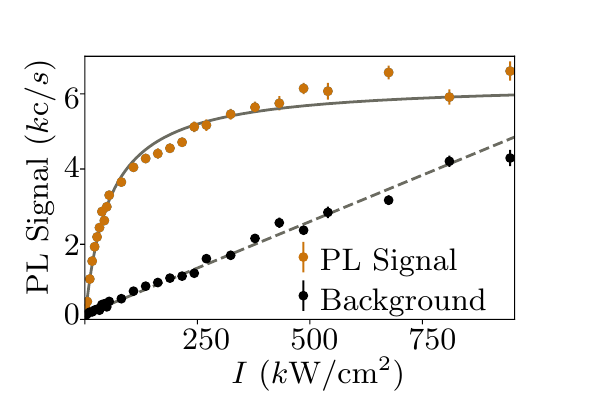}

\caption{PL-signal at different intensities $I$. The PL-signal is fitted by:
$S=S_{max}I/(I_{0}+I)$ with (6.3 $\pm$ 0.1) $k$cps and $I_{0}$=
(53$\pm$ 2) kW/cm$^{2}$, whereas the background increases linearly
with the slope m = (4.9 $\pm$ 0.1) counts cm$^{2}$/(kW s) and intercept
b = (130 $\pm$ 8) cps.\label{fig:Experimentally-obtained-PL-signa}}
\end{figure}

A suitable electronical excitation Intensity $I$ can be obtained
from the saturation behaviour of the PL-signal. In Fig.$\:$\ref{fig:Experimentally-obtained-PL-signa},
a maximum PL- signal of (6.3 $\pm$ 0.1) $k$cps can be obtained,
while the background increases linearly with increasing $I$. Therefore,
$I_{0}$ = (53 $\pm$ 2) kW/cm$^{2}$ is chosen for ODMR-measurements,
since the PL-signal is almost saturated while the background is relatively
small. Even if a high background is mostly compensated by the references
detection \textgreek{D}PL, it has the advantage that the autofocus
works more reliably at low background.

\section*{Appendix G: ODMR simulations}

\begin{figure}
\includegraphics[scale=1.5]{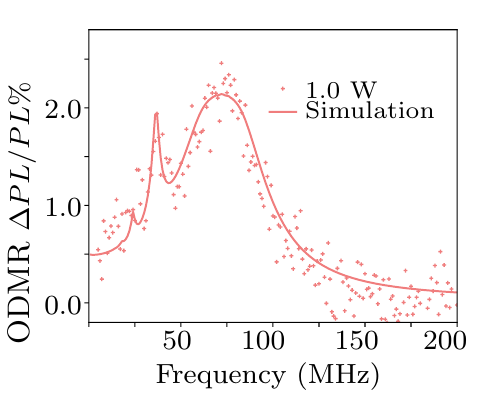}

\caption{Experimental and simulated ODMR plots for silicon vacancy with the
RF Hamiltonian of Eq.~\ref{eq:lindblad}.}

\label{odmrandsim}
\end{figure}

\begin{figure}
\includegraphics{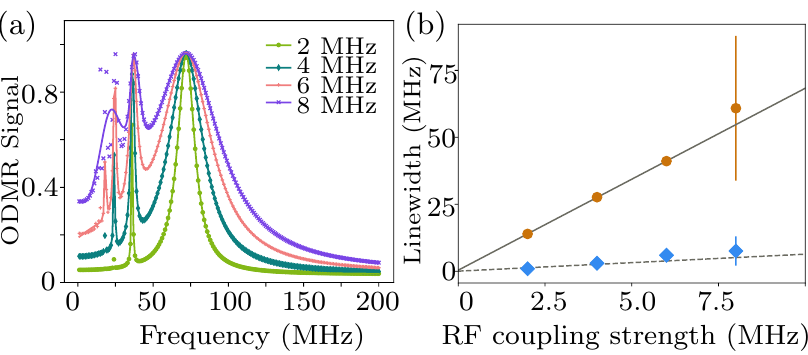}

\caption{(a) Simulated ODMR spectra at different RF strengths. (b) Linewidth
of 1-photon and 2-photon resonances vs. RF coupling strength.}

\label{sim_odmr}
\end{figure}

For the numerical simulation, we used the Lindblad master equation
\begin{eqnarray}
\frac{\partial\rho}{\partial t} & = & -2\pi i\:[{\cal H}_{t}(t),\rho]+\sum_{\alpha,\beta,\delta_{1,..,5}}L_{i}^{\dagger}.\rho.L_{i}-\frac{1}{2}\left\{ L_{i}^{\dagger}L_{i},\rho\right\} \label{eq:lindblad}
\end{eqnarray}
here ${\cal H}_{t}(t)={\cal H}+{\cal H}_{RF}(t)$, $L_{\alpha}=\sqrt{2}\left(\begin{array}{cccc}
0 & \sqrt{\gamma} & 0 & 0\\
\sqrt{\gamma} & 0 & \sqrt{\alpha} & 0\\
0 & \sqrt{\alpha} & 0 & \sqrt{\gamma}\\
0 & 0 & \sqrt{\gamma} & 0
\end{array}\right)\thickapprox\sqrt{2\alpha}\:S_{x}$ drives the spin-lattice relaxation process, $\gamma=3\alpha/4$,
$L_{\beta}=\sqrt{2\beta}\:S_{z}$ is the Lindblad operator for the
dephasing process, and $L_{\delta_{1,..,5}}$ are the Lindblad operators
for the optical pumping~\cite{singh-prr-2022}

$L_{\delta_{1}}=\sqrt{\delta}\left(\begin{array}{cccc}
0 & 0 & 0 & 0\\
1 & 0 & 0 & 0\\
0 & 0 & 0 & 0\\
0 & 0 & 0 & 0
\end{array}\right)$, $L_{\delta_{2}}=\sqrt{\delta}\left(\begin{array}{cccc}
0 & 0 & 0 & 0\\
0 & 0 & 0 & 1\\
0 & 0 & 0 & 0\\
0 & 0 & 0 & 0
\end{array}\right)$,

$L_{\delta_{3}}=\sqrt{\delta}\left(\begin{array}{cccc}
0 & 0 & 0 & 0\\
0 & 0 & 0 & 0\\
1 & 0 & 0 & 0\\
0 & 0 & 0 & 0
\end{array}\right)$, $L_{\delta_{4}}=\sqrt{\delta}\left(\begin{array}{cccc}
0 & 0 & 0 & 0\\
0 & 0 & 0 & 0\\
0 & 0 & 0 & 1\\
0 & 0 & 0 & 0
\end{array}\right)$,

$L_{\delta_{5}}=\sqrt{\delta}\left(\begin{array}{cccc}
0 & 0 & 0 & 0\\
0 & 0 & 1 & 0\\
0 & 1 & 0 & 0\\
0 & 0 & 0 & 0
\end{array}\right)$, and $\delta$ is the optical pumping rate~\cite{singh-prb-21}.

The interaction Hamiltonian between the RF field and the spins is

\[
{\cal H}_{RF}(t)=\Omega_{1}\text{cos}(2\ensuremath{\pi}\omega t)(S_{x}+S_{z}),
\]
where $\Omega_{1}$ = $g\mu_{B}B_{1}$ represents the strength of
the coupling to the RF field $B_{1}$ in frequency units and $\omega$
is the oscillation frequency of the field. Figure~\ref{odmrandsim}
shows the experimental and simulated ODMR spectra at 8 MHz RF coupling
strengths with $\alpha$ = 7 ms$^{-1}$, $\beta$ = 10 $\mu$s$^{-1}$,
and $\delta=$ 185 ms$^{-1}$; the simulated spectrum matches well
with the experimentally recorded ODMR at 1 W RF power. Since we are
using simple optical pumping model~\cite{singh-prb-21}, the amplitude
of the simulated spectrum is normalized to the maximal ODMR contrast
for the plot. For the simulated spectrum, we integrated Eq.~\eqref{eq:lindblad}
for 1.5 $\mu$s for every frequency. Additional details can be found
in our previous work~\cite{singh-prr-2022}. Figure~\ref{sim_odmr}
(a) shows the simulated ODMR spectra at different RF coupling strengths
with the same parameters used for simulating ODMR spectrum in Fig~\ref{odmrandsim}
except for the value of $\beta$, which we reduced here to 2.5 $\mu s^{-1}$
to better resolve the multi-photon peaks.

\begin{table}
\begin{tabular}{|c|c|c|}
\hline 
Peak & $LW_{0}$(MHz) & $a$\tabularnewline
\hline 
\hline 
1 & 0.34 $\pm$0.21 & 6.80 $\pm$0.05\tabularnewline
\hline 
2 & $\approx$0 & 0.63 $\pm$0.05\tabularnewline
\hline 
\end{tabular}

\caption{Obtained fitting parameters from Eq.~\eqref{eq:linewidth-1} for
the linewidth vs RF field strength.}

\label{simlinewidthpar}
\end{table}

In Fig~~\ref{sim_odmr}, with an RF coupling strength of 2 MHz,
we can see 1-photon, and 2-photon peaks at 72 MHz and 36 MHz and a
small signal of 3-photon peak at 24 MHz. At 8 MHz, the 5-, 4- and
3-photon peaks are also visible and close to each other, and therefore
difficult to resolve. The linewidth of the 1- and 2-photon peaks are
extracted by fitting the simulated ODMR to a sum of Lorentzians. Fig~\ref{sim_odmr}
(b) shows the resulting dependence of the linewidth $LW$ vs. the
square root of the RF power $\varLambda$ itogether with a fir to
the function

\begin{equation}
LW(P)=LW_{0}+a_{1}\,\Omega_{1}.\label{eq:linewidth-1}
\end{equation}
The obtained fitting parameters, intercept $LW_{0}$ and the slope
$a$ are given in Table~\eqref{simlinewidthpar}.

\section*{Appendix H: ODMR with different rf and laser powers}

\begin{figure}
\includegraphics{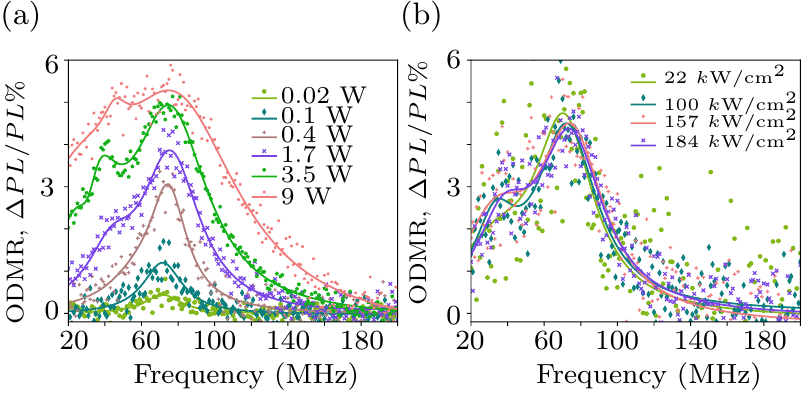}\caption{ODMR spectra recorded for one of the high contrast centers (out of
ten measured centers), (a) with different applied RF power and 184
kW/ cm\texttwosuperior{} laser intensity; (b) with different laser
intensities and 2.6 W RF power. The horizontal axis is the frequency
in MHz, and the vertical axis is the relative change of the PL\%.}

\label{odmr_dif_rf_pow}
\end{figure}

Figure~\ref{odmr_dif_rf_pow} (a) shows the ODMR signal recorded
for one of the high-contrast centers with different RF power, keeping
the laser intensity at 184 kW/ cm\texttwosuperior{} and (b) for different
Laser intensity at 2 W RF power. At 9 W RF power, the ODMR signal
contrast is 5.3 \%, and the laser intensity does not affect the ODMR
contrast significantly. In the course of recording these data, the
PL count of the center fell from 8 kcps to 4 kcps which impacts the
signal-to-noise ratio of the later recorded variable laser intensity
ODMR spectra.
\begin{acknowledgments}
This work was supported by the Deutsche Forschungsgemeinschaft in
the frame of the ICRC TRR 160 (Project No. C7). J.U.H. acknowledge
support from Swedish Research Council under VR grant No.2020-05444
and Knut and Alice Wallenberg Foundation (grant no. KAW 2018.0071).
J.U.H. and F.K. acknowledge support from the European Union through
the QuantERA grant InQuRe. F.K. acknowledges support from the German
ministry of education and research (BMBF, grant no. 16KIS1639K, 16KISQ013
and 13N16219).
\end{acknowledgments}

\end{document}